\documentclass[%
 reprint,
superscriptaddress,
 amsmath,amssymb,
 aps,
]{revtex4-2}
\usepackage[colorlinks=true,
linkcolor=blue,
urlcolor=blue,
citecolor=blue]{hyperref}
\usepackage{cancel}
\usepackage{graphicx}
\usepackage{dcolumn}
\usepackage{bm}
\usepackage{url}
\usepackage{bm}
\usepackage{bbm}   
\usepackage{verbatim}
\usepackage{stmaryrd}
\usepackage{amsthm}
\usepackage{amssymb}
\usepackage{array,multirow}
\usepackage{empheq}
\usepackage[usenames,dvipsnames]{color}
\usepackage[normalem]{ulem}
\usepackage{tikz}
\usepackage{algorithm}
\usepackage{algcompatible}
\usepackage{algpseudocode} 

\usepackage{soul}

\usepackage{transparent}

\theoremstyle{plain}

\theoremstyle{definition}

\theoremstyle{remark}

\newcommand{\kB}{k_\mathrm{B}}  
\newcommand{\ts}{t_\mathrm{s}}
\newcommand{\Vc}{V_\mathrm{c}}
\newcommand{\Vr}{V_\mathrm{r}}
\newcommand{\nul}{\nu_\mathrm{low}}
\newcommand{\nuh}{\nu_\mathrm{high}}

\begin{document}

\title{Energetic cost of feedback control}
\author{Jannik Ehrich}
\email{jehrich@sfu.ca}
\affiliation{Department of Physics, Simon Fraser University, Burnaby, British Columbia,V5A 1S6, Canada}
\affiliation{Department of Physics and Astronomy, University of Hawaii at M\=anoa, Honolulu, Hawaii 96822, USA}
	
\author{Susanne Still}
\email{sstill@hawaii.edu}
\affiliation{Department of Physics and Astronomy, University of Hawaii at M\=anoa, Honolulu, Hawaii 96822, USA}
	
\author{David A.\ Sivak}
\email{dsivak@sfu.ca}
\affiliation{Department of Physics, Simon Fraser University, Burnaby, British Columbia,V5A 1S6, Canada}
	
\begin{abstract}
Successful feedback control of small systems allows for the rectification of thermal fluctuations, converting them into useful energy; however, control itself requires work. This paper emphasizes the fact that the controller is a physical entity interacting with the feedback-controlled system. For a specifically designed class of controllers, reciprocal interactions become nonreciprocal due to large timescale separation, which considerably simplifies the situation. We introduce a minimally dissipative controller model, illustrating the findings using
a simple example.
We find that the work required to run the controller must at least compensate for the decrease in entropy due to the control operation.
\end{abstract}
	
\maketitle

\section{Introduction}
Originally inspired by the work of Maxwell~\cite{Maxwell1867,Leff2002_Maxwells} and Szilard~\cite{szilard1929_On_the}, numerous experiments within the last decade have demonstrated that feedback control of small systems can be used to rectify thermal fluctuations, turning them into useful energy~\cite{Serreli2007_Molecular,Bannerman2009_Single-photon,Toyabe2010_Experimental,Koski2014_Experimental_Obs,Koski2014_Experimental_Realiz,Koski2015_On-chip,Vidrighin2016_Photonic,Camati2016_Experimental,Chida2017_Power,Cottet2017_Observing,Paneru2018_Losless,Masuyama2018_Information-to-work,Naghiloo2018_Information,Admon2018_Experimental,Paneru2018_Optimal,Ribezzi2019_Large,Paneru2020_Efficiency,Saha2021_Maximizing,Saha2022_Bayesian}. However, to achieve the task of rectifying fluctuations, information about the controlled system is needed in order to determine the appropriate feedback actions. But acquiring, storing, and processing information costs energy. The resulting dissipation balances or exceeds the benefit gained from feedback control, in accordance with the second law~\cite{szilard1929_On_the,landauer1961irreversibility,bennett1982thermodynamics,Maruyama2009_Colloquium,Sagawa2009_Minimal,still2014lossy, Sandberg2014_Maximum,Horowitz2014_Second_law-like,Parrondo2015_Thermodynamics,Um2015_Total,Still2020_cost_benefit_memory, Still2021_Partially}. Understanding the trade-off is thus of great practical interest.

Theoretical work has analyzed far-from-equilibrium thermodynamic processes involving measurements with subsequent feedback, and situations with repeated feedback loops~\cite{Sagawa2010_Generalized,Cao2009_Thermodynamics,Ponmurugan2010_Generalized,Horowitz2010_Nonequilibrium,Sagawa2012_Nonequilibrium,Ehrich2017_Stochastic,crooks2019marginal}. Continuous-time implementations of feedback control have also been considered~\cite{Strasberg2013_Thermodynamics,Hartich2014_Stochastic,Horowitz2014_Info_Flow,Sandberg2014_Maximum,Horowitz2014_Second_law-like,Barato2017_Thermodynamic_cost,Ciliberto2020_Autonomous,Freitas2021_Characterizing,Ehrich2022_Energy}. 

We focus here on information engines that apply repeated feedback at discrete points in time by making a measurement of the system state, modifying a feedback potential, and letting the system relax in the potential until the next measurement. A plethora of experimentally realized information engines and theoretical models fit this paradigm. One key distinction is whether the relaxation time is so long that the system equilibrates between measurements~\cite{Toyabe2010_Experimental,Um2015_Total,Paneru2018_Losless,Paneru2020_Efficiency,Still2021_Partially}, or sufficiently short that subsequent measurements are correlated~\cite{Bauer2012_Efficiency,Schmitt2015_Molecular,Bechhoefer2015,Chida2017_Power,Admon2018_Experimental,Paneru2018_Optimal,Ribezzi2019_Large,Saha2021_Maximizing,Lucero2021_Maximal,Saha2022_Bayesian}.

Whenever the state of the controller depends on the measurement of the system, which is a random variable, the control itself becomes a cofluctuating random variable, i.e., the system and controller form a bi-variate stochastic process in which each system is influenced by the other. Energy flows between strongly coupled, cofluctuating systems contain, in general, both heat-like and work-like contributions~\cite{crooks2019marginal}; however, the situation may be simplified by applying specific design assumptions to the controller, which we pursue here.

A system controlled by a parameter that is adjusted using knowledge of the system's state was studied extensively~\cite{Sagawa2012_Fluctuation, Sagawa2014_Role}, and it was found that the entropy production of the controlled system is bounded by a difference in mutual information. This bound, however, does not immediately give insight into the minimum thermodynamic costs of the feedback-control operation. In particular, an open question is whether the difference in mutual information is actually the minimum realizable work required to run the controller. To illuminate this issue, we use a bottom-up approach, starting from a physical controller model to explicitly calculate the work required to achieve feedback control.
Our method differs from a previous approach~\cite{Horowitz2014_Second_law-like} which started with an abstract inequality and then added an interpretation. We confirm that the minimum work required for implementing feedback control by updating the controller according to a particular feedback rule is given by an information-theoretic quantity that can be related to the difference in mutual information between controller and system, before and after the controller update. We illustrate how a physically realizable controller can reach the minimum work. This approach allows us to derive an information-theoretic quantification of the cost of feedback control using a familiar expression for the work done on small fluctuating systems.

In this paper, we account for both the conversion of information to work \emph{and} the work required to record and react to the information.
Contrary to previous approaches that regard the feedback controller as \emph{external} to the system and thus require
a separate specification of a measurement process distinct from the system dynamics,
we assume here that
feedback-controlled system and controller {\em together} form an information engine. We emphasize the fact that the controller is a physical entity that can only interact with the feedback-controlled system via interaction potentials.
The interaction potentials are designed such that an external experimenter only supplies predetermined modifications of the potential to realize the desired feedback control, and hence, does not need to know the actual system state.
The resulting reciprocal interactions between feedback-controlled system and controller can then become effectively nonreciprocal~\cite{Loos2020_Irreversibility} in the limit of large timescale separation between the two component's dynamics.

Section~\ref{sec2} specifies our physical controller model and gives the lower bound on the work required to run this type of controller, together with a protocol that achieves it. The example
in Sec.~\ref{sec:example_feedback_cooling} illustrates the situation.

\section{Model of feedback controller}
\label{sec2}
We consider the joint time evolution of a feedback-controlled system $X$ and a controller $Z$. The system $X$ is assumed to be small and in contact with a thermal bath at temperature $T$, which results in stochastic dynamics. Let $x(t)$ and $z(t)$ be the respective states of system and controller at time $t$. As an example, consider overdamped dynamics for both the feedback-controlled system $X$ and the controller $Z$ which interact via a (conservative) potential-energy landscape. For such a setup, heat and work are readily calculated~\cite{Sekimoto1998_Langevin,Sekimoto2010_Stochastic,Seifert2012_Stochastic}.

We model information engines that employ repeated feedback, where measurement and feedback happen cyclically, with sampling period $\ts$. The time $\tau$ to measure system $X$ and update the controller based on this measurement is assumed to be much smaller than the time between sampling: $\tau \ll \ts$. In these engines the measurement is assumed to happen without back-action. The system state is measured at times $k\, \ts$, resulting in $x_k := x(k\ts)$, where $k =\{0,..., K\}$. Using this information, the controller updates to a new value $z_{k} := z(k\ts +\tau)$.

In the idealized case, the update time vanishes, $\tau \rightarrow 0$. In the following, we assume that during the time when the controller is updated, $X$ does not change. 

Between controller updates, the system changes dynamically, and the controller state is stable (in the sense that fluctuations are negligible), so we can write $z[(k+1)\ts] = z(k \ts+\tau) = z_{k}$.

During this time interval the energy of subsystem $X$ depends on the controller state. We consider the overdamped case, where only the potential energy affects dynamics. The relaxation potential $\Vr(x,z_{k})$ controls the dynamics of subsystem $X$ during the $k$th relaxation step, giving time-dependent potential energy $\Vr[x(t),z_{k}]$.

We assume alternating stability: the system does not change during the controller update, and the controller is stable during system relaxation. This can be achieved by fast controller updates and making the controller mobility sufficiently small during the relaxation steps, e.g., by making the controller sufficiently large.

The externally applied driving protocol thus modifies not only the interaction energy between $X$ and $Z$, but also changes the mobility $\nu_z(t)$ of the controller between a low mobility $\nul$ to keep it as unchanged as possible during the system relaxation step, and a high mobility $\nuh$ to rapidly update the controller during the control step.
Changing the controller mobility in this way and assuming a short controller update time $\tau$ ensures that the subsystems are alternatingly stable: Each subsystem's relaxation time is too long to react during the other subsystem's update, thus introducing an effective non-reciprocity even though the forces on both subsystems are derived from a joint potential.

\begin{figure}[t]
    \centering
	\includegraphics[width = 1\linewidth]{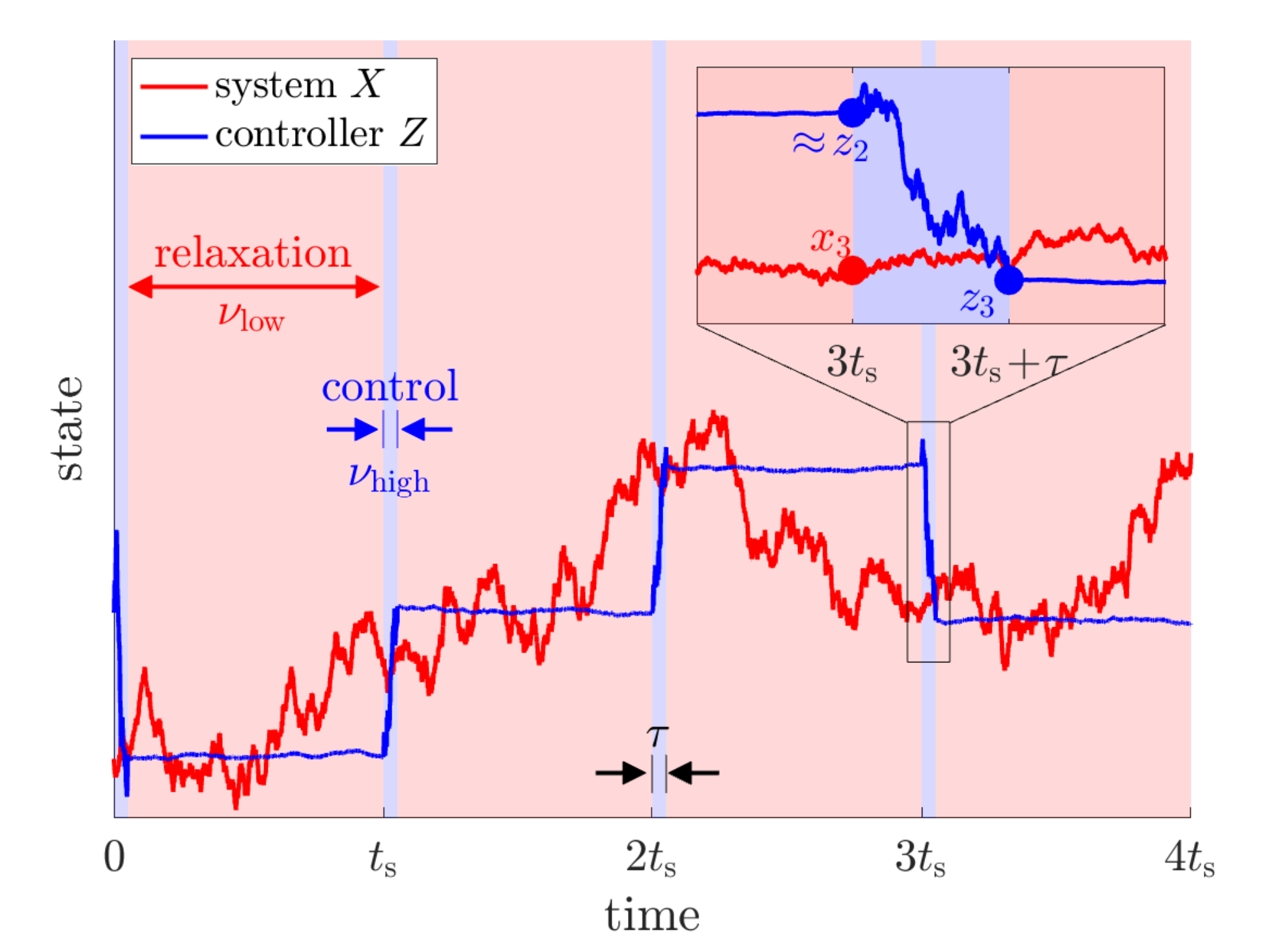}
	\caption{A process with repeated feedback. A nearly stable controller $Z$ with a very low mobility $\nul$ provides a trapping potential for a colloidal particle $X$. At times $k\ts, k=\{0,1,...\}$, the controller's mobility is increased to $\nuh$ and the controller is quickly recentered on the particle through a fast update taking time $\tau \ll \ts$. This process approximates an idealization in which the controller is stable during system relaxation and instantaneously updates its position at periodic intervals.}
	\label{fig:illustrative_feedback_process}
\end{figure}

Figure~\ref{fig:illustrative_feedback_process} illustrates the repeated-feedback process for a system that consists of a colloidal particle in a trap and a controller that periodically moves the trap center to the particle position (examined in detail in Sec.~\ref{sec:example_feedback_cooling}). During the relaxation step the trap position changes very little, but during the control step it quickly recenters on the particle, which has almost no time to respond.

\begin{figure}[b]
    \centering
	\includegraphics[width = 1\linewidth]{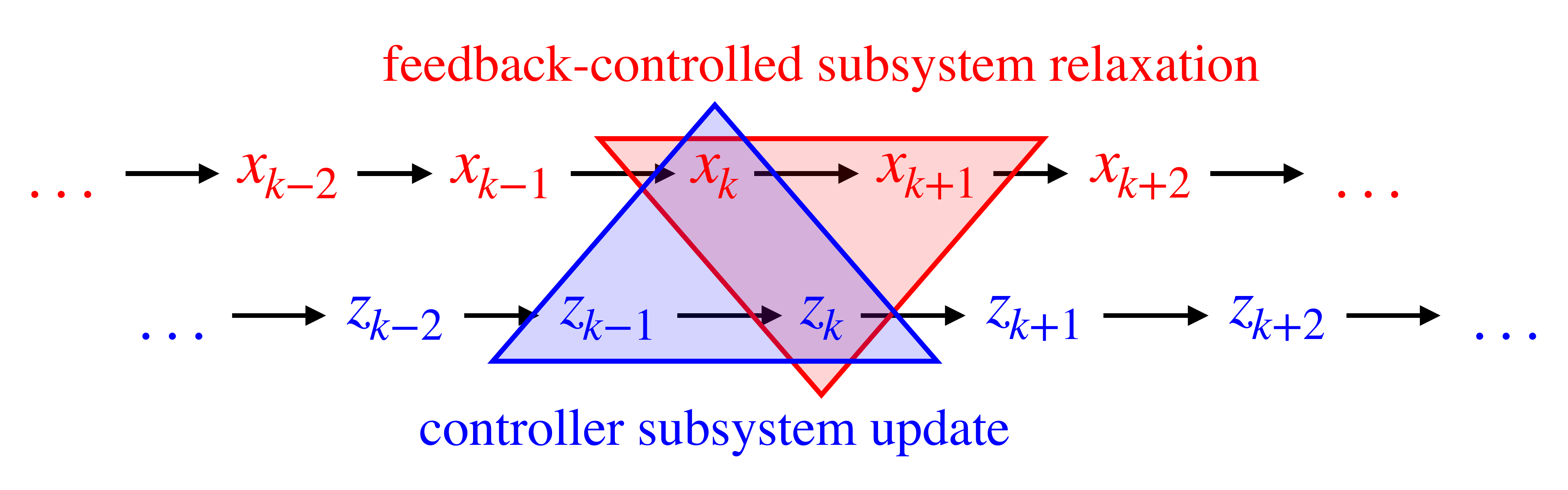}
	\caption{Temporal ordering of the discrete-time process resulting from assuming alternating stability. Updates of the controller state $z$ only happen in the control step, and updates of the state $x$ of the feedback-controlled system only occur in the relaxation step.}
	\label{fig:subsystem_updates}
\end{figure}

A discrete-time notation updates the timestep counter only once in each cycle with the temporal ordering: $x_k \rightarrow z_{k} \rightarrow$ timestep-counter update $\{ k \rightarrow k+1\}$ $\rightarrow x_{k+1} \rightarrow z_{k+1}$. Figure~\ref{fig:subsystem_updates} illustrates the temporal ordering of the discrete-time dynamics.

\subsection{Heat absorbed by feedback-controlled system} \label{sec:heat_fb_system}
The total potential-energy change over a cycle is $\Vr(x_{k+1},z_{k}) - \Vr(x_k,z_{k-1})$, which can be split into two contributions; $\Vr(x_{k+1},z_{k}) - \Vr(x_{k},z_{k})$ when the controller is fixed and $\Vr(x_{k},z_{k}) - \Vr(x_k,z_{k-1})$ when the feedback-controlled subsystem $X$ is fixed.

We assume that during the time period in which $X$ evolves dynamically under $V_r(x,z_{k})$ and changes its potential energy from $\Vr(x_{k},z_{k})$ to $\Vr(x_{k+1},z_{k})$, the experimenter does no work on the joint $X$-$Z$ system, so only heat
\begin{equation}
    q^X_k := \Vr(x_{k+1},z_{k}) - \Vr(x_{k},z_{k}) \label{eq:heat_X}
\end{equation}
is exchanged with the environment. (Our convention is that heat absorbed by and work done on the system are positive.)

\subsection{Apparent work}
If the controller subsystem $Z$ was an external control parameter (i.e., not considered part of the system) rather than a cofluctuating subsystem, then $q^X_k$ would be the only heat exchanged with the environment, and the work done on $X$ by $Z$ would be
\begin{equation}
    w^{\rm app}_k := \Vr\left(x_{k},z_{k} \right) - \Vr\left( x_{k},z_{k-1} \right)\,. \label{eq:idealized_feedback_work}
\end{equation}
Thus, if the experimenter supposed in their accounting that there was no feedback control, then they would assume that this ought to be the work done on subsystem $X$. But in reality, there is feedback control, and this is not the actual work. For that reason, we refer to it as ``apparent'' work.

Note that if there was no feedback control, then subsystem $X$, driven by control parameter $Z$, would have the nonequilibrium (``generalized'') free energy~\cite{Gaveau1997_general_framework,Gaveau2008_Work_and_power,Esposito2011_NonequilibriumFreeEnergy,SivakCrooks_PRL12,Still2012_Thermodynamics} associated with the conditional distribution: $F[X|Z]:=~\langle \Vr(x,z) \rangle_{p(x,z)} - \kB T H[X|Z]$, for Boltzmann constant $\kB$ and conditional entropy~\cite[Chapter 2.2]{Cover2006_Elements} $H[X|Z]~:=~- \langle \ln\left[p(x|z)\right] \rangle_{p(x,z)}$.

The average work dissipated during a controller update $z_{k-1} \to z_k$ at fixed $x_k$ would then be
\begin{subequations}
\begin{align}
    \left\langle w^\mathrm{app}_k \right\rangle - \Delta F_k^Z[X|Z]  &= \kB T \left( H[X_{k}|Z_{k}] - H[X_{k}|Z_{k-1}] \right) \\
    &= \kB T \left( I[X_{k};Z_{k-1}] -  I[X_{k};Z_{k}] \right) \,, \label{eq:app_EP}
\end{align}
\end{subequations}
with mutual information~\cite[Chapter 2.3]{Cover2006_Elements} $I[X;Z]:=~H[X] - H[X|Z]$.
Therefore,
the lower bound on
total average
dissipated work would be proportional to the instantaneous nonpredictive information which the $X$-system keeps about the control signal $Z$~\cite{Still2012_Thermodynamics}.

\subsection{Total work done on the joint system}
Crucially, in the case of the cofluctuating controller subsystem $Z$, $w^{\rm app}_k$ is just one part of the work done on the joint system. The work to update the controller from $z_{k-1}$ to $z_{k}$, which can be considerable, also needs to be accounted for.

The task of the controller is to rectify thermal fluctuations of the system, thereby converting input heat into output work.
To achieve this, the controller itself is externally controlled
by an experimenter; however, this external control is limited to modulating the coupling potential in a predetermined way such that the controller implements the desired feedback. In particular, the external experimenter need not know the system state $x_k$ or a measurement thereof. (Section~\ref{sec:min_work_prot} illustrates how this can be achieved.) Nonetheless, the experimenter needs to supply the work required
to change the controller's state.

We model manipulations of the controller with a time-dependent control potential $\Vc(x,z;t)$, thereby filling in the detailed temporal development of the controller between $z_{k-1}$ and $z_{k}$. Our treatment is thus a more specialized version than the treatment in \cite{crooks2019marginal}. The control potential steers the controller from state $z_{k-1}$ to state $z_{k}$ at constant subsystem state $x_{k}$. To achieve a controller update sufficiently fast to hold the subsystem fixed, we switch the controller mobility from $\nul$ to $\nuh$ before the controller update and back to $\nul$ at its end. In cycle $k$, the controller update steps are: 
\begin{enumerate}
    \item[(i)] At time $k\ts$: instantaneous switch from the relaxation potential to the start of the control potential, $\Vr(x,z) \to \Vc[x,z;k\ts]$, and from low to high controller mobility, $\nul \to \nuh$.
    \item[(ii)] Between $t = k\ts$ and $ t =k\ts + \tau$: continuous manipulation of the control potential $\Vc(x,z;t)$, thereby bringing the controller to the new value, $z_{k}$.
    \item[(iii)] At time $kt_s+\tau$: instantaneous switch to new relaxation potential, $\Vc[x,z;k\ts+\tau] \to \Vr(x,z)$, and from high to low controller mobility, $\nuh \to \nul$.
\end{enumerate}
To account for all steps, the following energy changes contribute to the work done on the joint $X$-$Z$ system~\cite{Sekimoto1998_Langevin,Sekimoto2010_Stochastic,Jarzynski2011_Equalities,Seifert2012_Stochastic,VandenBroeck2015_Ensemble}: 
\begin{align}
    w_k &= \Vc\left( x_{k},z_{k-1},k\ts\right) - V_\mathrm{r}(x_{k},z_{k-1})\nonumber\\
    &\qquad+ \int\limits_{k\,\ts}^{k\,\ts+\tau} dt\,\frac{\partial \Vc(x_{k},z;t)}{\partial t} \bigg|_{z(t)} \nonumber \\
    &\qquad+ V_\mathrm{r}(x_{k},z_{k}) - \Vc \left(x_{k},z_{k};k\ts+\tau\right) \,.\label{eq:total_work}  
\end{align}
We define work in excess of apparent work as the ``additional work'' necessary to achieve the controller update: 
\begin{equation}
    w^{\rm add}_k := w_k - w^{\rm app}_k\,. \label{eq:def_add_work}
\end{equation}
The average work done on the joint $X$-$Z$ system over the entire protocol is $W = \left\langle \sum_{k=0}^{K} w_k \right\rangle$. Including the heat flow $q_k^Z$ while $X$ is fixed and only $Z$ changes [step (ii) above], the total work and heat over a cycle are $w_k + q_k = w^{\rm app}_k + w^{\rm add}_k + q^X_k + q^Z_k$, where $q_k := q^X_k + q^Z_k$, and the first law dictates
\begin{equation}
\label{eq:FirstLawCycle}
    \Vr(x_{k+1},z_{k}) - \Vr(x_{k},z_{k-1}) = w^{\rm app}_k + w^{\rm add}_k + q^X_k + q^Z_k ~.
\end{equation}
The definitions of $q^X_k$ and $w^{\rm app}_k$, however, yield 
\begin{equation}
\label{eq:SumWQ}
    w^{\rm app}_k + q^X_k = \Vr(x_{k+1},z_{k}) - \Vr(x_{k},z_{k-1}) \ ,
\end{equation}
illustrating the purpose of an information engine: converting input heat into output work. Combining Eqs.~\eqref{eq:FirstLawCycle} and \eqref{eq:SumWQ} reveals that all additional work to effect the controller change must be dissipated as heat: $w^{\rm add}_k = -q^Z_k$. In the following we will address how this additional work can be minimized.

\subsection{Minimum additional work} \label{sec:minimum_add_work}
The nonequilibrium free energy of the joint system is $F[X,Z]:= \langle V(x,z) \rangle_{p(x,z)} - \kB T H[X,Z]$, for joint entropy $H[X,Z] = - \langle \ln\left[p(x,z)\right] \rangle_{p(x,z)}$. The free-energy change over a cycle is then, using the first law over a cycle~\eqref{eq:FirstLawCycle}:
\begin{subequations}
\begin{align}
    \Delta F_k &:= F[X_{k+1},Z_{k}] - F[X_k, Z_{k-1}] \\
    &= \langle w_k \rangle + \langle q^X_k + q^Z_k \rangle - \kB T \Delta H^{X,Z}_k\,.
\end{align}
\end{subequations}
We split the entropy change into contributions from changing $X$ and changing $Z$:
\begin{subequations}
\begin{align}
    \Delta H^{X,Z}_k &:= H[X_{k+1},Z_{k}] - H[X_k,Z_{k-1}] \\
    &= H[X_{k+1}|Z_{k}] - H[X_k|Z_{k}] \nonumber\\
    &\qquad+ H[Z_{k}|X_{k}] - H[Z_{k-1}|X_{k}] \,, \label{eq:condEnts}
\end{align}
\end{subequations}
where line~\eqref{eq:condEnts} adds and subtracts $H[X_k|Z_k]$ and uses the chain rule~\cite[Chapter\ 2.5]{Cover2006_Elements} for joint, conditional, and marginal entropies, $H[A,B] = H[A|B] + H[B]$.

We analogously split the free energy into $\Delta F_k = \Delta F^Z_k + \Delta F^X_k$, summing the contribution during adjustment of the controller $Z$,
\begin{align}
    \Delta F_k^Z &:= \left\langle w_k\right\rangle + \left\langle q^Z_k \right\rangle - \kB T (H[Z_{k}|X_{k}] - H[Z_{k-1}|X_{k}]) \label{eq:DFZ_end}\,,
\end{align}
and the contribution while the subsystem $X$ evolves in the relaxation potential $\Vr$,
\begin{equation}    
    \Delta F_k^X := \left\langle q^X_k\right\rangle - \kB T (H[X_{k+1}|Z_{k}] - H[X_k|Z_{k}]) \label{eq:DFX_2}\,.
\end{equation}
The second law implies that $\langle w_k \rangle - \Delta F_k^Z  \geq 0$, and thus
\begin{equation}
    \langle w^{\rm add}_k \rangle = - \langle q^{Z}_k \rangle \geq \kB T \left( H[Z_{k-1}|X_{k}] - H[Z_{k}|X_{k}] \right)\,. \label{LBwck}
\end{equation}

In summary, the additional work needed to adjust the controller has to at least compensate for the decrease in conditional entropy due to the controller update. The remaining uncertainty, quantified by $H[Z_{k}|X_{k}]$, measures the precision with which the controller adjusts its state $z_k$ to the feedback-controlled subsystem state $x_k$. On the other hand, $H[Z_{k-1}|X_{k}]$ corresponds to the precision of the controller's anticipation of the next subsystem state. Higher-precision controller updates impose larger minimum thermodynamic costs; conversely, higher-precision controller anticipation reduces minimum thermodynamic costs.

The entropy production by the joint $X$-$Z$ system over one time step, $\Delta H^{X,Z}_k - \beta \langle q_k \rangle := H[X_{k+1},Z_k] - H[X_k,Z_{k-1}] - \beta \langle q^X_k + q^Z_k\rangle$, where $\beta = 1/k_\mathrm{B}T$, can thus be bound by the sum of the change in entropy, $\Delta H^{X}_k := H[X_{k+1}] - H[X_k]$, the heat produced when $X$ changes, and the instantaneous nonpredictive information that the controller $Z$ retains about the feedback-controlled system $X$, $I_{\rm nonpred, k}^{Z \rightarrow X}:= I[Z_k;X_k] - I[Z_{k};X_{k+1}]$ ~\cite{Still2012_Thermodynamics}. This quantity is equal to the negative information flow, $I^X_{\rm flow, k} := I[X_{k+1};Z_k] - I[X_k;Z_k]$~\cite{Horowitz2014_Info_Flow}. Thus we have:
\begin{subequations}
\label{eq:bound_interpr}
\begin{align}
\Delta H^{X,Z}_k - \beta \langle q_k \rangle 
&\geq  \Delta H^{X}_k - \beta \langle q^X_k \rangle + I_{\rm nonpred, k}^{Z \rightarrow X} \label{eq:bound_interpr_1} \\
&\quad = \Delta  H^{X}_k - \beta \langle q^X_k \rangle - I^X_{\rm flow,k}~.
\end{align}
\end{subequations}
Less dissipation is required for a controller-update rule that (on average) performs better at predictively inferring the next state of the system, and thus captures less instantaneous nonpredictive information for the same amount of memory. 

This is also reflected in the fact that average additional work required to run the controller, summed over an entire experiment, $W^{\rm add}: = \sum_{k=0}^{K-1} w^{\rm add}_k$, can not be less than the total instantaneous nonpredictive information that the controller state keeps about the signal which is causing it to change (feedback-controlled subsystem $X$), $I_{\rm nonpred}^{Z \rightarrow X}:=\sum_{k=0}^{K-1} \left( I[Z_k;X_k] - I[Z_{k};X_{k+1}] \right)$, minus the overall increase in precision, $\Delta H^{Z|X} := H[Z_K|X_K] - H[Z_0|X_0]$:
\begin{equation}
    \beta W^{\rm add} \geq I_{\rm nonpred}^{Z \rightarrow X} - \Delta H^{Z|X}~. \label{WLB}
\end{equation}
Bounds on the apparent work $\langle w^\mathrm{app}_k\rangle$ extractable using feedback control, that are similar to the RHS of~\eqref{eq:bound_interpr} and \eqref{WLB}, have been found~\cite{Sagawa2012_Fluctuation,Sagawa2014_Role} without an explicit controller model. We find here that the RHS of Eqs.~\eqref{eq:bound_interpr} and \eqref{WLB} provides the minimum additional work required for control, as carried out by a real-world, physically implemented controller.
Importantly, our analysis does not require an external observer's measurement. Instead, control is carried out mechanistically by a time-dependent modification of the coupling potential, implementing the effect of a measurement and subsequent feedback by an external observer.
In the next subsection we illustrate how our explicit controller architecture achieves 
the
minimum
additional work.

The controller-update rule could be optimized to be predictive and thus minimize dissipation, as proposed in \cite{Still2020_cost_benefit_memory}; however, in most practical applications, a controller-update rule $p_{\rm c}(z_{k}|x_{k})$ is chosen (often heuristically) by the experimenter. Let us thus assume for the remainder of this paper that the update rule is given. This probabilistic rule could describe, e.g., a noisy measurement and the subsequent controller reaction to it. For the simple case of $z$ drawn from a Gaussian $p_{\rm c}(z|x)$ with mean $x$ and standard deviation $\sigma$, the intuition for $H[Z_{k}|X_{k}] = \ln \sigma + \ln \sqrt{2\pi} +1/2$ is straightforward: sloppier adjustments result in a wider distribution (larger $\sigma$) and hence, larger $H[Z_{k}|X_{k}]$.

\subsection{Protocols minimizing additional work}
\label{sec:min_work_prot}
Together with the given $X$-dynamics, characterized by $p(x_{k+1}|x_{k}, z_{k})$, and an initial condition $p(x_0)$, the distribution $p(z_{k-1}| x_{k}) = p(x_{k}, z_{k-1}) / \sum_{z_{k-1}} p(x_{k}, z_{k-1})$ is computed recursively via $p(x_{k}, z_{k-1}) = \sum_{x_{k-1}} p(x_{k}|x_{k-1}, z_{k-1}) p_{\rm c}(z_{k-1}|x_{k-1}) p(x_{k-1})$.

With both initial and final distributions fixed by the given controller-update rule and system dynamics, we seek to minimize the average controller work. The lower bound of Eq.~\eqref{LBwck} is reached when $\langle w^{\rm add}_k \rangle/ \kB T = H[Z_{k-1}|X_{k}] - H[Z_{k}|X_{k}]$. This can be achieved by choosing~\cite{Takara2010_Generalization}: 
\begin{subequations}
\begin{align}
    \Vc(x, z ;k\ts) &= -\kB T \ln p(z_{k-1}=z|x_{k}=x) \label{stepA}\\
    \Vc(x,z;k\ts +\tau) &= -\kB T \ln p(z_{k}=z | x_{k}=x) \,, \label{stepB}
\end{align}
\end{subequations}
and {\it quasistatically} changing $\Vc$ between $t = k\ts$ and $t = k\ts+\tau$, which is ensured by a sufficiently large controller mobility $\nu \gg \tau^{-1}$. Similar protocols have been used to minimize thermodynamic costs when copying polymers~\cite{Ouldridge2017_Fundamental}.

The limits of interest, which lead to an idealized feedback process, are: $\nul \to 0$ makes the controller stable during the relaxation step; $\tau \to 0$ makes the control step sufficiently short that the system is effectively immobile; $\nuh \to \infty$ with fixed large $\nuh \times \tau \gg 1$ ensures that the controller-update protocol realizes the quasistatic step, and thus approaches the minimum-work implementation.

This idealization in terms of a concurrent double-timescale separation achieves an effectively quasistatic update of the controller that is simultaneously instantaneous from the point of view of the system.

The bound on the (average) minimum additional work required for feedback control, Eqs.~\eqref{LBwck} and \eqref{WLB}, together with the description of a protocol for a physical controller that reaches the bound, are our main results. These results highlight that feedback control needs to be understood and analyzed as being carried out by a physical system.

To summarize: Instead of assuming that an ethereal external observer makes a measurement and executes feedback on a feedback-controlled system, we highlight the fact that the controller is a physical system that is coupled to the feedback-controlled system via carefully designed interaction potentials. Reproducing the alternating stability found in many example measurement-feedback processes requires a clear separation of time scales between controller and feedback-controlled system. In our model, control is achieved by reversibly changing the controller's state, which is realized through a time-dependent control potential and a high controller mobility. Because the system state is stable during this (fast) update, the system state selects the distribution for the next controller state as the conditional equilibrium distribution of the control potential. Standard energetic considerations permit calculation of the minimum work needed for this update, which is bounded by the free-energy change during the update, leading to an information-theoretic lower bound.

%\section{Example: Feedback cooling}
\section{Example feedback process}
\label{sec:example_feedback_cooling}
In the previous section we have found a bound on the minimum work required for feedback control. Here, we illustrate this finding by studying
in detail a model of a simple measurement-feedback process in the idealized limit. We describe the model (Sec.~\ref{sec:model_description}) and solve for its dynamics (Sec.~\ref{sec:model_dynamics}). In Sec.~\ref{sec:feedback_work} we calculate the apparent work without recourse to any model of how the control is achieved. Next, in Sec.~\ref{sec:control_work} we use the arguments of Sec.~\ref{sec:minimum_add_work} to give a lower bound on the additional work needed to achieve the desired control. In Sec.~\ref{sec:explicit_control_model}, we relax the idealization by allowing for a finite-but-short feedback time $\tau$. We then give an explicit model for the controller's dynamics realizing the minimum-work protocol described in Sec.~\ref{sec:min_work_prot}. We numerically simulate the entire process of feedback and relaxation and calculate the work done on the joint system to verify that the lower bound on the additional work is reached in the idealized limit.

\subsection{Model description}\label{sec:model_description}
Consider the position $x$ of a Brownian particle obeying overdamped dynamics with Stokes friction coefficient $\gamma$. The particle diffuses in a harmonic trapping potential with stiffness $\kappa$. The controller's state $z$ is the position of the trap center. The particle's dynamics, given a controller state $z$, evolve according to the Langevin equation
\begin{equation}
    \dot x = -\left(x-z\right) + \sqrt{2}\,\xi(t)\,, \label{eq:Langevin_model}
\end{equation}
where $\xi(t)$ denotes Gaussian white noise with zero mean and covariance $\left\langle \xi(t)\xi(t')\right\rangle = \delta(t-t')$, and we rescaled lengths by the standard deviation $\sqrt{\kB T/\kappa}$ of the equilibrium distribution and times by the particle's relaxation time $\gamma/\kappa$ in the trap. The relaxation potential is therefore $\Vr(x,z) = \frac{1}{2}(x-z)^2$, and energy is measured in units of $\kB T$.

Feedback consists of the controller periodically measuring the particle at times $t = k\,t_\mathrm{s}$, $(k= 1,..., K)$, and then updating its own state to the measurement outcome, thereby recentering the trap at the measurement. For simplicity, we assume that the measurement has a Gaussian error with mean zero and standard deviation $\sigma$, such that $z_{k} = x_k +\sigma \hat\eta$, where $\hat\eta$ is a zero-mean Gaussian random variable with unit variance. When $\sigma$ is sufficiently small, the controller ``tracks'' the particle and extracts energy from the particle's fluctuations. The resulting information engine resembles the model studied in \cite{Paneru2020_Efficiency}, but does not reset the trap position to zero after each time step.

\subsection{Model dynamics}\label{sec:model_dynamics}
Integrating Eq.~\eqref{eq:Langevin_model} over one time step gives the conditional distribution~\cite[Sec. 4.5.4]{Gardiner2004_Handbook}
\begin{equation}
   p\left(x_{k+1}|x_k,z_{k}\right) = \mathcal{N}\left[x_{k+1};z_{k}+(x_k-z_{k}) e^{-t_\mathrm{s}}, 1 - e^{-2t_\mathrm{s}} \right]\,, \label{eq:prop_x}
\end{equation}
where $\mathcal{N}(x;\mu,c)$ denotes a Gaussian distribution of $x$ with mean $\mu$ and variance $c$. The update rule for the controller is
\begin{equation}
   p_\mathrm{c}\left(z_{k}|x_k\right) = \mathcal{N}\left(z_{k}; x_k, \sigma^2 \right)\,. \label{eq:meas_distribution}
\end{equation}

Assuming that initially the particle is at $x_0=0$ and the controller is distributed around it according to Eq.~\eqref{eq:meas_distribution}, the time evolution of $x$ is
\begin{subequations}
\begin{align}
    p(x_{k+1}|x_k) &= \int dz_{k} \, p\left(x_{k+1}|x_k,z_{k}\right)\,p_\mathrm{c}\left(z_{k}|x_k\right)\\
    &= \mathcal{N}(x_{k+1};x_k,\Delta c_{xx})\,,
\end{align}
\end{subequations}
where 
\begin{equation}
    \Delta c_{xx} = \sigma^2 \left(1-e^{-t_\mathrm{s}} \right)^2 + 1 - e^{-2 t_\mathrm{s}}
\end{equation}
is the increment in particle variance from one time step to the next. Therefore, the marginal particle distribution is
\begin{equation}
    p(x_k) =\mathcal{N}\left(x_k;0, k \Delta c_{xx}\right)\,.
\end{equation}

Because the controller periodically recenters the trap around the fluctuating particle, on timescales longer than the feedback time $\ts$ the dynamics of the particle correspond to free diffusion with effective diffusion coefficient $\Delta c_{xx}/2\ts$. The joint distribution of particle and ensuing controller state is
\begin{equation}
    p(x_k,z_k) = \mathcal{N}\left[ \begin{pmatrix} x_k\\z_k
    \end{pmatrix}; \begin{pmatrix} 0 \\ 0
    \end{pmatrix},\mathbf{C}^{x_k}_{z_k}
    \right]\,, \label{eq:joint_prob}
\end{equation}
where $\mathcal{N}\left[\mathbf{z};\boldsymbol{\mu},\mathbf{C}\right]$ is a multivariate Gaussian distribution of $\mathbf{z}$ with mean vector $\boldsymbol \mu$ and covariance matrix $\mathbf C$. Here,
\begin{align}
    \mathbf{C}^{x_k}_{z_{k}} := 
    \begin{pmatrix} k\Delta c_{xx} & k\Delta c_{xx}\\
    k\Delta c_{xx} & k\Delta c_{xx} + \sigma^2
\end{pmatrix}\,. \label{eq:cov_matrix}
\end{align}

Due to the operation of the feedback, if $\sigma < 1$, the measurement is sufficiently precise that the particle distribution 
\begin{equation}
    p(x_k|z_{k}) = \mathcal{N}(x_k;z_{k},\sigma^2)
\end{equation}
with respect to the trap center, is narrower than the corresponding equilibrium distribution, which can be interpreted as a lower effective temperature of the particle. The model therefore realizes an overdamped version of \emph{feedback cooling}, which usually refers to feedback forces leading to velocity distributions being narrower than the equilibrium Maxwell-Boltzmann distribution in underdamped systems~\cite{Kim2007_Fluctuation, Poggio2007_Feedback}.

\subsection{Apparent work}
\label{sec:feedback_work}
The feedback operation changes the internal energy of the joint system, which is interpreted as \emph{apparent work} done on the system. Following Eq.~\eqref{eq:idealized_feedback_work}, the average apparent work per control step (in scaled units) is
\begin{subequations}
\begin{align}
    \left\langle w^\mathrm{app}_k\right\rangle
    &= \Big\langle \frac{1}{2}(x_{k}-z_{k})^2 - \frac{1}{2} (x_{k}-z_{k-1})^2 \Big\rangle\\
    &= \frac{1}{2}\Big\langle z_{k}^2 - z_{k-1}^2 - 2 x_{k} z_{k} + 2 x_{k} z_{k-1} \Big\rangle \,. \label{eq:work_model_end}
\end{align}
\end{subequations}
To calculate this work explicitly, we require the joint distribution of the particle and the previous controller state, which, using Eqs.~\eqref{eq:prop_x} and \eqref{eq:joint_prob}, becomes
\begin{subequations}
\begin{align}
    p(x_{k},z_{k-1}) &= \int dx_{k-1}\, p_x(x_{k}|x_{k-1},z_{k-1}) p(x_{k-1},z_{k-1})\\
    &=\mathcal{N}\left[ \begin{pmatrix} x_{k}\\z_{k-1}
    \end{pmatrix} ; \begin{pmatrix} 0 \\ 0
    \end{pmatrix}, \mathbf{C}^{x_{k}}_{z_{k-1}} \right] \,, \label{eq:joint_prob_different_times}
\end{align}
\end{subequations}
for the covariance matrix of particle position and controller position after the relaxation ($X$ update) and before the control step ($Z$ update),
\begin{align}
    &\mathbf{C}^{x_{k}}_{z_{k-1}}= \\
    &\begin{pmatrix}
    k\Delta c_{xx} & (k\!-\!1) \Delta c_{xx}\! + \sigma^2\!\left( 1 \!-\! e^{-t_\mathrm{s}}\right) \\
    (k\!-\!1) \Delta c_{xx} +  \sigma^2\!\left(1\! -\! e^{-t_\mathrm{s}}\right) & (k\!-\!1)\,\Delta c_{xx} + \sigma^2
    \end{pmatrix}\,. \label{eq:cov_matrix_diff_times} \nonumber
\end{align}
Then, the apparent work in Eq.~\eqref{eq:work_model_end} reads
\begin{equation}
    \left\langle w^\mathrm{app}_k\right\rangle = \frac{1}{2}\left( 1- e^{-2t_\mathrm{s}} \right) \left( \sigma^2 -1 \right)\,. \label{eq:work_model_explicit}
\end{equation}
When $\sigma < 1$, this is negative, indicating work extraction from thermal fluctuations.

\begin{figure}[tb]
    \centering
    \includegraphics[width = 1 \linewidth]{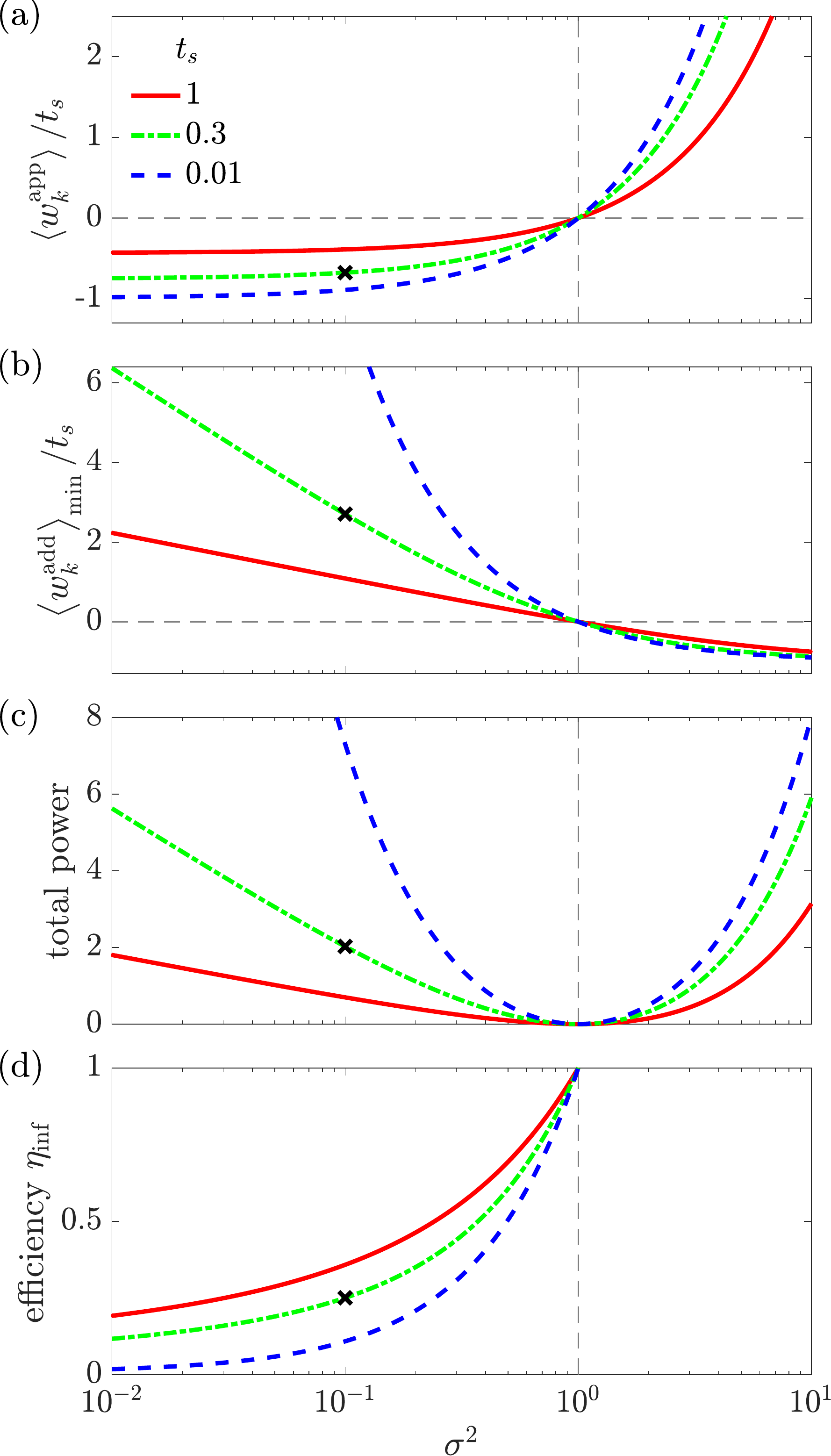}
    \caption{Performance of the feedback-cooling engine. (a) Rate of apparent work~\eqref{eq:work_model_explicit} as a function of the measurement variance $\sigma^2$ for different sampling times $t_\mathrm{s}$. Negative work indicates work extraction. (b) Rate of minimum additional work~\eqref{eq:non_pred_model}. (c) Sum of rates of apparent and additional work, showing that apparent work is more than compensated by additional work. (d) Information efficiency~\eqref{eq:def_info_efficiency} defined as the ratio of extracted work (benefit) to minimum additional work (costs). Black crosses: parameter combination used in Fig.~\ref{fig:explicit_model_sim}.}
    \label{fig:model_performance}
\end{figure}

Figure~\ref{fig:model_performance}(a) shows the rate of work extraction, $\left\langle w^\mathrm{app}_k\right\rangle/t_\mathrm{s}$. The smallest input work rate corresponds to the maximum extracted work rate, which is $\lim\limits_{\sigma \to 0, t_\mathrm{s} \to 0} \left\langle w^\mathrm{app}_k \right\rangle/t_\mathrm{s}=-1$, reached for very accurate ($\sigma \to 0$) and frequent ($t_\mathrm{s} \to 0$) measurements. Regardless of sampling period $t_\mathrm{s}$, at $\sigma=1$ the measurement uncertainty equals the position variation in the trap, such that feedback does not change the width of the position distribution, so $\left\langle w^\mathrm{app}_k\right\rangle=0$.

We also calculate the heat flowing into subsystem $X$ during its relaxation,
\begin{subequations}
\begin{align}
    \left\langle q^X_k \right\rangle &= \left\langle \Vr(x_{k+1},z_{k}) - \Vr(x_{k},z_{k}) \right\rangle\\   
    &= \frac{1}{2}\left\langle x_{k+1}^2 - x_{k}^2 - 2 x_{k+1} z_{k} + 2 x_{k} z_{k}  \right\rangle\\
    &= -\frac{1}{2}\left( 1- e^{-2t_\mathrm{s}} \right) \left( \sigma^2 -1 \right) \label{eq:heat_model}\\
    &= -\left\langle w^\mathrm{app}_k \right\rangle\,, \label{eq:heat_equal_work_model}
\end{align}
\end{subequations}
illustrating that the engine extracts apparent work from heat that flows into the system during the relaxation step.

\subsection{Additional work and efficiency}\label{sec:control_work}
Following Sec.~\ref{sec:minimum_add_work}, we calculate the lower bound~\eqref{LBwck} on the additional work from the joint probabilities of subsystem and controller [Eqs.~\eqref{eq:joint_prob} and \eqref{eq:joint_prob_different_times}]. The lengthy general expression simplifies in the limit $k\to \infty$ to
\begin{subequations}
\begin{align}
    \left\langle w^{\rm add}_k\right\rangle_\mathrm{min} &=H[Z_{k-1}|X_{k}] - H[Z_{k}|X_{k}]\\
    &= H[X_{k},Z_{k-1}] - H[X_{k},Z_{k}] \\
    &= \frac{1}{2}\ln\frac{\left|\mathbf C^{x_{k}}_{z_{k-1}}\right|}{\left|\mathbf C^{x_{k}}_{z_{k}}\right|}\\
    &\overset{k \to \infty}{\longrightarrow} \frac{1}{2}\ln\frac{1 + \left( \sigma^2 -1 \right) e^{-2 t_\mathrm{s}}}{\sigma^2}\,, \label{eq:non_pred_model}
\end{align}
\end{subequations}
where $|\cdot|$ is the determinant.

Figure~\ref{fig:model_performance}(b) shows the minimum rate of average additional work. Comparing with Fig.~\ref{fig:model_performance}(a), negative apparent work is accompanied by positive additional work, and vice versa. Figure~\ref{fig:model_performance}(c) verifies that the average total work is nonnegative. The negative apparent work is thus more than compensated by costs incurred in running the controller. The total work only vanishes at $\sigma=1$, because before and after the feedback the particle distribution in the trap is the same equilibrium distribution, and hence feedback does not change the free energy. For $\sigma > 1$ the roles of the feedback-controlled system and controller reverse in some respects, and positive apparent work is converted into negative additional work.

Figure~\ref{fig:model_performance}(b) illustrates that the costs of running the controller increase with increasing measurement accuracy ($\sigma \to 0$). Moreover, the costs increase with feedback frequency ($t_\mathrm{s} \to 0$): Frequent, accurate measurements are costly.

This finding is illustrated by the information efficiency
\cite{Bauer2012_Efficiency,Sandberg2014_Maximum}
defined as benefit (extracted work) relative to costs (additional work),
\begin{subequations}
\begin{align}
    \eta_\mathrm{inf} &:= \frac{-\left\langle w^\mathrm{app}_k \right\rangle}{\left\langle w^{\rm add}_k\right\rangle_\mathrm{min}} \\
    &= \frac{-\langle w_k^\mathrm{app} \rangle}{k_B T (H[Z_{k-1}|X_k] - H[Z_k|X_k])} \ , \label{eq:def_info_efficiency}
\end{align}
\end{subequations}
shown in Fig.~\ref{fig:model_performance}(d) for the $k \to \infty$ limit. This measure of information efficiency is maximized at vanishing output power. Faster feedback and more accurate measurements reduce possible efficiency.

\subsection{Explicit physical controller model} \label{sec:explicit_control_model}
Here, we present an explicit model of a controller that realizes the minimum average additional work. We assume that the controller state $z$ is described by a Langevin equation such that the dynamics of feedback-controlled system $X$ and controller $Z$ evolve according to the coupled Langevin equations
\begin{subequations}
\label{eq:joint_Langevin_model}
\begin{align}
    \dot x &= -\partial_x V(x,z;t) + \sqrt{2}\,\xi_x(t) \label{eq:joint_Langevin_model_x}\\
    \dot z &= -\nu_z\partial_z V(x,z;t) + \sqrt{2\nu_z}\, \xi_z(t)\,, 
    \label{eq:joint_Langevin_model_z}
\end{align}
\end{subequations}
where $\xi_x(t)$ and $\xi_z(t)$ are uncorrelated Gaussian white noises and $\nu_z$ is the piecewise-constant controller mobility switching between a large value $\nuh$ during the control step and a small value $\nul$ during the relaxation step to achieve the joint system's alternating stability as described in Sec.~\ref{sec2}. 

Although challenging, at least conceptually the dynamics of this joint system could be realized experimentally by a Brownian particle with tunable anisotropy in a two-dimensional potential-energy landscape that could be generated, e.g., by virtual potentials using a feedback trap~\cite{Gavrilov2016_Feedback,Kumar2018_Nanoscale}.

The relaxation potential is the quadratic $\Vr(x,z) = \frac{1}{2}(x-z)^2$. The control potential $\Vc(x,z;t)$ quasistatically and reversibly carries the controller from the previous conditional controller distribution $p(z_{k-1}=z|x_{k}=x;t) \propto \exp\left[-\Vc(x,z;k\ts) \right]$ at time $k\ts$ to the next conditional controller distribution $p(z_{k}=z|x_{k}=x) \propto \exp\left[-\Vc(x,z;k\ts+\tau) \right]$ at time $k\ts+\tau$. (Recall that the control step is sufficiently short, $\tau \ll 1$, that the particle does not move, $x(t) \equiv x_{k}$ for $t \in [k\ts,k\ts + \tau]$.) 

To this end, we dynamically change the control potential according to
\begin{equation}
    \Vc(x,z;t) = \frac{1}{2} \kappa(t) (x-z)^2\,,
\end{equation}
for time-dependent trap stiffness
\begin{equation}
    \kappa(t) = \left[\left(\sigma^2-1\right) \left(1-e^{-2\ts}\right)\frac{(t\; \mathrm{mod}\; k\ts) -\tau}{\tau} + \sigma^2\right]^{-1}\; 
\end{equation}
which linearly interpolates between 
\begin{subequations}
\begin{align}
   \kappa(k\ts) &= \mathrm{Var}^{-1}\left[z_{k-1}|x_{k} \right]\\
   &= \left[\left\langle \left(z_{k-1} - \left\langle z_{k-1} \right\rangle_{p(z_{k-1}|x_{k})} \right)^2 \right\rangle_{p(z_{k-1}|x_k)}\right]^{-1} \\
   &= \left[\left( \sigma^2 -1 \right) e^{-2\ts} + 1\right]^{-1} \quad \mathrm{for}\; k \to \infty \,, \label{eq:stiffness_final_pot}
\end{align}
\end{subequations}
calculated from Eq.~\eqref{eq:joint_prob_different_times}, and $\kappa(k\ts + \tau) = \mathrm{Var}^{-1}\left[z_{k}|x_{k} \right] = \sigma^{-2}$.

If all timescales are sufficiently separated ($\nuh^{-1} \ll \tau \ll \ts \ll \nul^{-1}$), then the controller update is effectively instantaneous, compared to the system dynamics and the feedback loop time $\ts$, but it also is effectively infinitely slow, compared to controller dynamics during the controller-update step.

We simulate the process by numerically integrating the coupled Langevin equations [\eqref{eq:joint_Langevin_model_x}, \eqref{eq:joint_Langevin_model_z}]. Figure~\ref{fig:illustrative_feedback_process}, which served to illustrate a general process with repeated feedback, in fact depicts a trajectory from this very process with parameters $\ts=9.5\cdot 10^{-2}$, $\tau=5\cdot10^{-3}$, $\nul=3\cdot 10^{-3}$, and $\nuh=8$ (for easy visual interpretation, we deliberately chose modest timescale separations).

In each timestep, the additional work~\eqref{eq:def_add_work} is
\begin{align}
    w^{\rm add}_k &= \Vc(x_{k},z_{k-1};k\ts) - \Vc(x_{k},z_{k};k\ts+\tau)\nonumber\\
    &\qquad+ \int\limits_{k\ts}^{k\ts+\tau} dt\,\frac{\partial \Vc(x_{k},z;t)}{\partial t} \bigg|_{z(t)}\,. \label{eq:model_control_work_integral}
\end{align}

\begin{figure}[htb]
    \centering
    \includegraphics[width = 1 \linewidth]{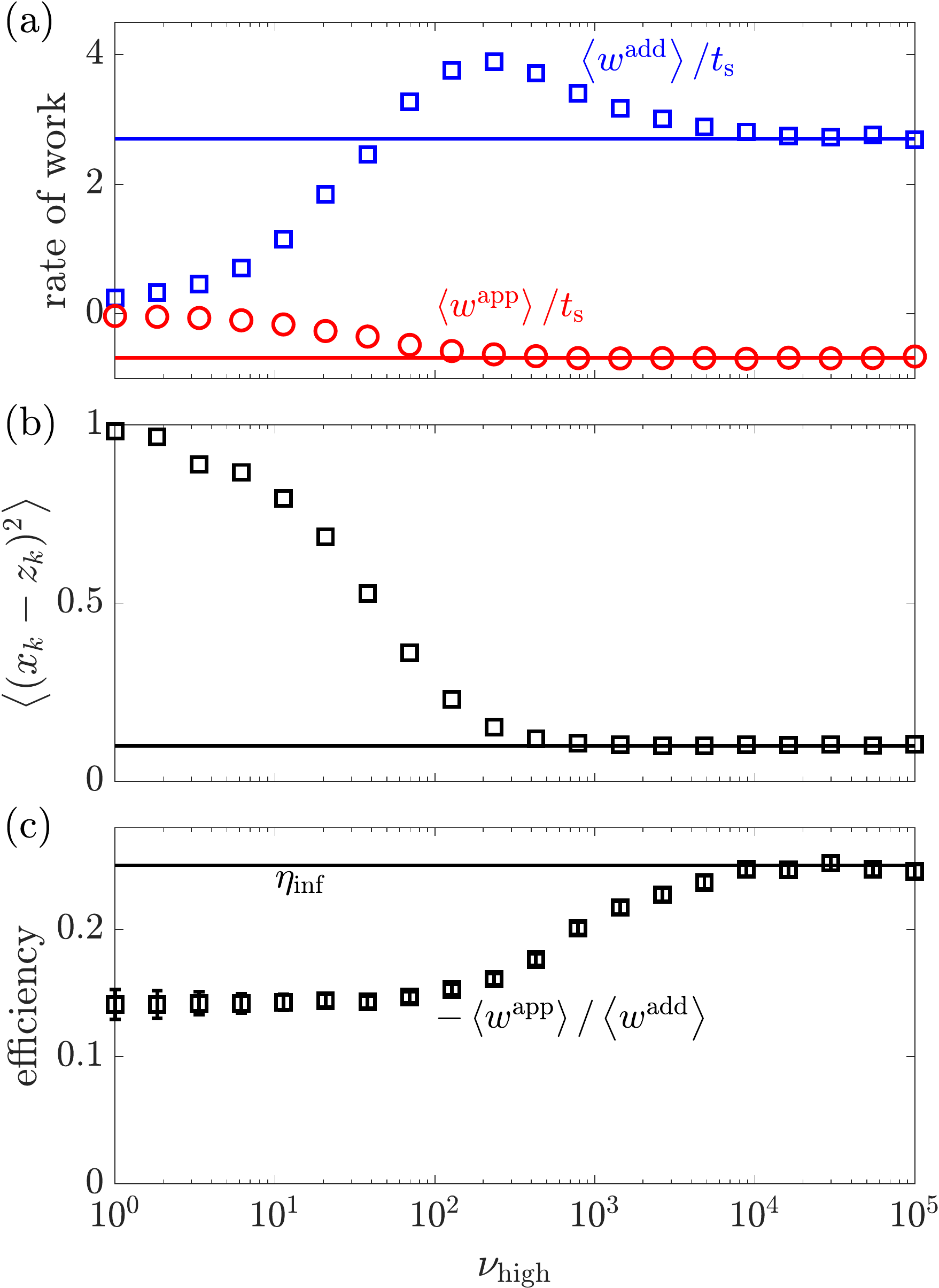}
    \caption{Performance of explicit controller implementation. (a) Rate of apparent work~\eqref{eq:idealized_feedback_work} and additional work~\eqref{eq:model_control_work_integral} as functions of controller mobility $\nuh$ during feedback, compared to respective model predictions~\eqref{eq:work_model_explicit} and \eqref{eq:non_pred_model} for $\nuh \to \infty$. Symbols are averages from $K=10^4$ timesteps, and standard errors of the mean are smaller than the symbol size. (b) Variance $\langle (x_k - z_k)^2 \rangle$ of the controller around the particle after the controller update. (c) Efficiency computed by dividing average apparent work (benefit) by average additional work (costs). Symbols are simulations and the solid line is information efficiency~\eqref{eq:def_info_efficiency}. Errorbars show standard errors of the mean. Simulation results are obtained by numerically integrating Eqs.~\eqref{eq:joint_Langevin_model_x}~and~\eqref{eq:joint_Langevin_model_z} with time step $dt=10^{-7}$ for feedback time $\ts=0.3$, measurement variance $\sigma^2=0.1$ (the parameters marked in Fig.~\ref{fig:model_performance}), controller-update time $\tau = 10^{-3}$, and controller mobility $\nul = 10^{-2}$ during the relaxation step.}
    \label{fig:explicit_model_sim}
\end{figure}

Figure~\ref{fig:explicit_model_sim}(a) shows the resulting average work contributions as a function of the controller mobility $\nuh$ during the control step. For large mobility ($\nuh \gtrsim 2\cdot 10^2$), the controller achieves its task to track the particle: Figure~\ref{fig:explicit_model_sim}(b) shows that the variance $\langle (x_k - z_k)^2\rangle$ of the controller around the particle after the control step matches the measurement variance $\sigma^2$. Consequently, the apparent work $\left\langle w^\mathrm{app}_k\right\rangle$ matches Eq.~\eqref{eq:work_model_explicit}.

For low $\nuh \lesssim 10^2$, the apparent work increases to zero and the additional work decreases to zero, because the controller state does not change appreciably during the control step, hence the joint system remains close to equilibrium and little work is done or extracted.

A controller mobility $\nuh \gtrsim 10^4$ is required for the apparent work to converge to the predicted value for $\nuh \to \infty$, which is negative (work is extracted), and for the additional work $\left\langle w^{\rm add}_k\right\rangle$ to achieve the lower bound given by~\eqref{eq:non_pred_model}. As $\nuh$ is decreased, changes in the control potential become too fast for the controller to track. Consequently, the control step is no longer quasistatic and the controller distribution lags behind its instantaneous equilibrium distribution, causing dissipation that results in greater additional work. 

To bound the resulting peak in additional work [Fig.~\ref{fig:explicit_model_sim}(a)], let us consider a ``worst-case'' estimate of the additional work needed to nonreversibly adjust the controller. Consider instantaneously setting the control to the desired final control potential $\Vc(x,z;k\ts +~\tau)$ with stiffness $1/\sigma^2$ and then letting the controller relax to its new equilibrium distribution. The work equals the heat released during the controller's relaxation, which can be bound using the relative variance $\mathrm{Var}\left[z_{k-1}|x_{k} \right]$ before the control step~\eqref{eq:stiffness_final_pot}. We obtain the estimate $\langle w_k^\mathrm{add}\rangle/\ts= -\langle q_k^Z \rangle/\ts = -1/(2\sigma^2)\times\left\{\sigma^2 - \mathrm{Var}\left[z_{k-1}|x_{k} \right]  \right\}/\ts \approx 8.2$. The maximum of the additional work in Fig.~\ref{fig:explicit_model_sim}(a) does not reach this value, because the mobility $\nuh$ is sufficiently high for the optimized process to harness some of the controller's relaxation dynamics, thus making the control step less costly than instantaneous switching.

Figure~\ref{fig:explicit_model_sim}(c) shows that the efficiency $-\left\langle w^\mathrm{app}_k \right\rangle/\left\langle w^{\rm add}_k \right\rangle$ increases with increasing $\nuh$, and is limited by $\eta_\mathrm{inf}$. The additional work $\left\langle w^{\rm add}_k \right\rangle$ falling below the minimum additional work $\langle w^{\rm add}_k\rangle_\mathrm{min}$ [Eq.~\eqref{eq:non_pred_model}] for small $\nuh$ in Fig.~\ref{fig:explicit_model_sim}(a) does not indicate higher efficiency because the lower additional work is more than compensated by lower extracted work.

\section{Discussion}
In this paper, we gave an expression for the minimum additional work as a function of a given feedback rule $p_\mathrm{c}(z_k|x_k)$. In some scenarios one might be interested in optimizing a control rule to maximize the total average work the engine produces~\cite{Still2020_cost_benefit_memory}, or other criteria. Together with the minimization of additional work we have pursued here, there could thus be a second optimization varying the feedback rule. Alternatively, both optimizations could be carried out together.

The study of the example system in Sec.~\ref{sec:explicit_control_model} shows that surprisingly large timescale separations are needed to achieve effectively instantaneous yet reversible control. If the controller mobility cannot exceed some maximal value, then control requires some minimum duration. The minimum-work protocol is then an \emph{optimal-transport} process, which has been studied in the context of finite-time thermodynamics
%~\cite{Aurell2011_Optimal,Aurell2012_Refined,Proesmans2020_Finite-time,Nakazato2021_Geometrical}.
\cite{Benamou2000_A_computational,Aurell2011_Optimal,Aurell2012_Refined,Proesmans2020_Finite-time,Nakazato2021_Geometrical,Chennakesavalu2021_Probing,Chennakesavalu2022_A_unified,Vu2022_Thermodynamic,Blaber2022_Optimal}
Finding the additional cost due to fast control would be an extension of the approach presented here.

Achieving the necessary timescale separation for alternating stability requires varying the mobility of the controller. Whether this requires additional thermodynamic costs is a matter of practical concern as it depends on the controller implementation. For example, if an electronic memory is used as the controller, then one may be able to raise and lower energy barriers between the controller's discrete states, thereby drastically changing mobility at vanishing extra costs.

We consider a static relaxation potential $\Vr(x,z)$. Although illustrative and simple to treat mathematically, this setup is not optimal: To harness all information gathered by the controller, a specifically designed, time-dependent feedback potential is required. Such a process can be made feedback-reversible~\cite{Horowitz2011_Thermodynamic_reversibility}. In our setup, the nonequilibrium relaxation dynamics of feedback-controlled subsystem $X$ in a static potential always cause entropy production, even if the distinct control step is perfectly reversible, as assumed here. This is not a severe limitation because a simple modification could make the relaxation potential time-dependent, $\Vr(x,z;t)$, allowing for feedback protocols that dissipate less heat and hence extract more work.

In our controller-update step, the control potential is specified as an evolving function of time. This may not be practical; a simpler but worse-performing alternative would set the desired final control potential $\Vc(x,z;k\ts +~\tau)$ at the beginning of the control step and rely on the large mobility of the controller to achieve relaxation to the correct final distribution. Such a protocol is easier to implement but does not take advantage of the controller relaxation during the update, so requires much higher additional work as explained in Sec.~\ref{sec:explicit_control_model}.

In contrast to other work on repeated-feedback processes~\cite{Cao2009_Thermodynamics,Horowitz2010_Nonequilibrium,Sagawa2012_Nonequilibrium}, our approach does not lead to transfer entropy~\cite{Schreiber2000_Measuring} or conditional mutual information as lower bounds for control cost. Our approach uses a controller-update rule that does {\it not} depend on the last controller state. A recursive update rule could be used instead: $p_\mathrm{c}(z_{k}|x_{k},z_{k-1})$, for which the next controller state $z_{k}$ would depend on the current controller state  $z_{k-1}$, which contains degrees of freedom storing memory of past measurements that are carried over to the next controller state. Recursive update rules which lead to lower dissipation are linked to learning and data-compression algorithms of the generalized information bottleneck class~\cite{Still_2014Entropy, Still_2023Entropy, Still_2009EPL}.

Modeling the controller as a physical system allowed us to identify the minimum work required to achieve the desired control through information processing and feedback in contrast to other approaches~\cite{Sagawa2010_Generalized,Ponmurugan2010_Generalized,Horowitz2010_Nonequilibrium,Sagawa2012_Nonequilibrium,Sagawa2012_Fluctuation,Sagawa2014_Role,Potts2018_Detailed} that only bound the extractable work achievable through feedback control without direct relation to the energetic cost of information processing. Having an explicit model of the controller's dynamics and its coupling to the feedback-controlled system alleviates interpretational ambiguities about the controller's operation, as can be found in, e.g., \cite{Horowitz2014_Second_law-like}.
The lower bound on the work necessary to update the controller can also be used to analyze operational costs of more complex information engines.

The utility of our setup is also reflected in the fact that there is no ambiguity about the time-reverse of a thermodynamic process with feedback. At first glance, the time-reverse of a feedback process might seem acausal, with effect (a specific control action) preceding cause (a measurement of the system state). Consequently, it has been common practice, when deriving fluctuation relations and second-law-like inequalities with information, to consider a reverse process that randomly picks a specific control protocol from the ensemble of forward control protocols and executes it in reverse without feedback~\cite{Sagawa2010_Generalized,Ponmurugan2010_Generalized,Horowitz2010_Nonequilibrium,Sagawa2012_Nonequilibrium}, where approaches have differed in whether measurements are made in the reverse process and whether some post-selection of the resulting trajectories is needed~\cite{Potts2018_Detailed}. Using, as we did, an exact specification of the potential and the controller mobility as a function of time, makes the time-reversed process transparently determined: it simply consists of executing the control on the joint system in reverse. With the initial condition of the reverse process starting from the final distribution of the forward process, entropy production and fluctuation theorems then follow straightforwardly~\cite{crooks2019marginal}.

\section{Conclusion}
In this paper we investigated information engines that employ repeated feedback, paying attention to the fact that the controller that realizes the desired feedback rule is a physical entity coupled to the feedback-controlled system via physical interaction potentials. We explicitly accounted for work needed to realize the prescribed controller dynamics. The average additional work needed to carry out the desired control cannot be less than the reduction in entropy it achieves. 

Our work highlights the fact that feedback control, including measurement, computation, and erasure of information needed to run an information engine, can be achieved \emph{mechanistically}. In our model these processes are completely internal to the joint system formed by a controller and feedback-controlled subsystem. The experimenter only supplies the scheduled modifications of the control potential and controller mobility and is not involved in any measurement or decision making.\\

The code for simulation and generating the plots in this paper can be found in Ref.~\cite{code}.

\begin{acknowledgements}
We thank John Bechhoefer, Gavin Crooks, Dorian Daimer, Joseph N.\ E.\ Lucero, Tushar K.\ Saha, and Rob Shaw for fruitful discussions. This research was supported by the Foundational Questions Institute, a donor-advised fund of the Silicon Valley Community Foundation, grant FQXi-IAF19-02 (all authors) and grant FQXi-RFP-1820 co-sponsored with the Fetzer Franklin Fund (S.S.). Additional support was from a Natural Sciences and Engineering Research Council of Canada (NSERC) Discovery Grant (D.A.S.) and a Tier-II Canada Research Chair (D.A.S.).
\end{acknowledgements}

\bibliography{refs_info_engine_costs}

\end{document}